# Probabilistic Scheduling of Dynamic I/O Requests via Application Clustering for Burst-Buffers Equipped HPC

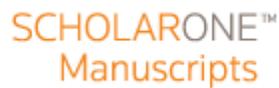

RESEARCH ARTICLE

# Probabilistic Scheduling of Dynamic I/O Requests via Application Clustering for Burst-Buffers Equipped HPC

Benbo Zha | Hong Shen

[1] School of computer science and engineering, Sun Yat-sen university, Guangzhou, China

**Correspondence**
Hong Shen, School of Computer Science and Engineering, Sun Yat-sen University, Guangzhou, China.
Email: shenh3@mail.sysu.edu.cn

**Summary**

Burst-Buffering is a promising storage solution that introduces an intermediate high-throughput storage buffer layer to mitigate the I/O bottleneck problem that the current High-Performance Computing (HPC) platforms suffer. The existing Markov-Chain based probabilistic I/O scheduling utilizes the load state of Burst-Buffers and the periodic characteristics of applications to reduce I/O congestion due to the limited capacity of Burst-Buffers. However, this probabilistic approach requires consistent I/O characteristics of applications, including similar I/O duration and long application length, in order to obtain an accurate I/O load estimation. These consistency conditions do not often hold in realistic situations.

In this paper, we propose a generic framework of **d**ynamic **p**robabilistic I/O **s**cheduling based on **a**pplication **c**lustering (DPSAC) to make applications meet the consistency requirements. According to the I/O phrase length of each application, our scheme first deploys a one-dimensional K-means clustering algorithm to cluster the applications into clusters. Next, it calculates the expected workload of each cluster through the probabilistic model of applications and then partitions the Burst-Buffers proportionally. Then, to handle dynamic changes (join and exit) of applications, it updates the clusters based on a heuristic strategy. Finally, it applies the probabilistic I/O scheduling, which is based on the distribution of application workload and the state of Burst-Buffers, to schedule I/O for all the concurrent applications to mitigate I/O congestion. The simulation results on synthetic data show that our DPSAC is effective and efficient.

## 1 | INTRODUCTION

High-performance computing platforms, particularly supercomputers, which provide enormous computing power and massive storage capacity, offer key infrastructures to accelerate modern scientific discovery through large-scale modeling and numerical simulation. Its applied fields are very extensive, such as high-energy physics, aerospace, bio-pharmacy, manufacturing, etc.[1] With the ever-increasing applications of HPC, the I/O capability of an HPC platform faces serious challenges in handling massive data processing tasks, known as the I/O bottleneck problem. The limitations of transferring data within HPC platforms severally restrict the implementation of data-intensive applications and the scale of applications.

There are three main reasons that are creating and exacerbating the current situation[2]. First, the data processing demands are increased rapidly because of very-large-scale scientific simulation experiments and Big-Data applications. Second, the growing gap between HPC computing nodes and the underlying storage components is becoming more significant as the development of computing techniques is faster than storage. Third, the fault-tolerance methods, such as Check-pointing/Restart strategy, are widely adopted by current HPC systems due to the decreasing mean time between failures (MTBF).

In order to mitigate the I/O bottleneck problem, many research efforts have been made, from designing different system architectures to optimizing application data layouts[3]. Two common research lines, Burst-Buffering and I/O scheduling, are closely related to our work. Burst-Buffering integrates a tier of high-throughput small-size storage devices, such as SSDs or NVRAMs, between the compute nodes and the underlying Parallel File System (PFS). By using the Burst-Buffers, HPC applications can absorb experienced I/O bursts (mainly write bursts). The architecture of Burst-Buffers that can be centralized or distributed was discussed in[4,5]. Some Burst-Buffer file systems have been designed in[6,7] to manage these storage resources uniformly. Some strategies for planing the capacity of Burst-Buffers have been studied in[8]. I/O scheduling is another common method to relieve the I/O bottleneck by reordering the I/O requests of concurrent applications and arranging each application's I/O rates at different intervals. Studies on I/O scheduling for HPC systems[9,10,11,12] are quite extensive, but relatively few on I/O scheduling for HPC platforms equipped with Burst-Buffers. Studies are often conducted from the viewpoint of data layout[13,14,15,16] and application coordination[17,18].

Because most applications running on HPC have obvious quasi-periodicity on I/O behavior, which alternates between computation and I/O transfer phases[19], many studies take this application characteristic into account. I/O scheduling for periodic applications on HPC was proposed[20,21]. The work[2] offers a probabilistic model for application to estimate the best size of Burst-Buffers. In our previous works[22,23], we combined the traditional I/O scheduling with Burst-Buffering to take advantage of the state of Burst-Buffers that is important to the performance of SSD due to Write Amplification[24]. Markov-Chain-based probabilistic I/O scheduling is proposed as well as two optimization techniques, which reduce repeated computation and the number of emptying, to mitigate I/O contention effectively by recognizing the degree of congestion on burst buffers. However, the underlying probabilistic model has an assumption that all applications have a similar length of I/O phrase and sufficiently long execution time[2].

These consistency conditions restrict the scope of the mentioned approaches' applicability in practical scenarios where applications concurrently running on an HPC platform may have different lengths and I/O characteristics. In this work, we propose a generic **d**ynamic **p**robabilistic I/O **s**cheduling framework based on **a**pplication **c**lustering (DPSAC), which clusters applications with a similar length of I/O phrase to satisfy the consistency condition. The framework includes four main components: Aggregator, Partitioner, Updater, I/O Scheduler. Aggregator deploys a one-dimensional K-means (1-D K-means) clustering algorithm to cluster applications according to their I/O lengths. Partitioner calculates the workload of each cluster through the probabilistic model of applications and partitions the burst buffers proportionally. Updater updates the configuration to cope with dynamic changes of applications via thresholding and online k-means updating. Finally, according to the framework's configuration, Scheduler decides the order of I/O transfers. The simulations conducted on synthetic data show that our framework can bypass the consistency condition.

The main contributions of this work are summarized below:

- We propose a generic probabilistic I/O scheduling framework based on application clustering (DPSAC) that enables applications in practical scenarios to meet the consistency requirements.
- We propose a cluster updating method based a heuristic strategy to handle dynamic changes (join and exit) of applications and promise performance via a more accurate estimation of application workload.
- We conduct extensive simulations on synthetic data based on real application characteristics that show the superiority of our proposed method against other methods.

The rest of this paper is organized as follows. In section 2, we described the related work on burst I/O buffering and I/O scheduling. The platform and application models, the objectives of I/O scheduling, and the motivation of our work are introduced in section 3. Section 4 describes our proposed generic probabilistic I/O scheduling framework detailedly. In section 5, we show the simulation and the results in detail. Section 6 concludes the paper.

## 2 | RELATED WORK

The I/O bottleneck problem in HPC becomes increasingly challenging with the increasing speed gap between computing and storage systems and the emerging of new demands on data processing and large-scale simulation. Numerous research efforts have been made to mitigate this problem. We review the most related work from three perspectives in this section.

### 2.1 | Burst-Buffering

As a promising technology, Burst-Buffering introduces a high-bandwidth, low-latency intermediate layer between computing and storage parts on HPC to absorb I/O bursts and mitigate I/O congestion[4]. There exist two implementations of the Burst-Buffers architecture[5]. Centralized strategy deploys the buffers on I/O nodes between the compute nodes and PFS. The DDN IME[25] and Cray Datawarp[26] apply this architecture to supercomputers. The distributed strategy allocates buffers to compute nodes[27] with the specific goals of fault tolerance.

To use the Burst-Buffers efficiently, Burst-Buffer file systems were designed in[6,7] to manage these storage resources uniformly. Strategies for planning the capacity of Burst-Buffers had been studied in[8]. A proactive draining method on Burst-Buffers is proposed in[16] to minimize I/O provisioning requirements. To reduce the effect of SSD's *write amplification*[24], A lazy strategy to empty Burst-Buffers was proposed in[2]. Our work is motivated by its probabilistic application model.

### 2.2 | I/O Scheduling on HPC

To mitigate I/O contention or achieve other optimizing targets, I/O scheduling is another standard technology that rearranges I/O requests or assigns different bandwidths for different applications. It can be designed on different layers of the storage hierarchy[28]. This work focuses on the PFS level, which studies coordinating I/O requests from one application or many applications.

For intra-application I/O scheduling, I/O requests come from many processes of one application. To utilize the data layout on disk, Collective I/O[29] applied a two-stage strategy to improve the application's I/O performance. To combine the topology of computing nodes, a topology-aware data aggregation method is provided in[30] to reduce the communication conflict. For across-application I/O scheduling, it coordinates the I/O requests from different applications. In[31], Dorier et al. proposed a scheduling scheme for two applications to reduce I/O interference. For more generic scenarios, I/O scheduling for many applications has been studied in[11,20] to increase system efficiency and decrease application dilation. The machine learning based method to extract I/O characteristics from logs is also presented in[9]. Another I/O coordinating technique in[32] is introduced to reorganize the I/O requests to utilize the spatial locality.

However, studies on I/O scheduling for Burst-Buffers equipped HPC are relatively few. In[14,15], the proposed SSDup distinguishes sequential and random requests to partition the underlying SSD into two identical half. Liang et al. in[13] balance the number of processes on each SSD server to reduce I/O contention. I/O interference on Burst-buffers is investigated in[33,17]. In[18], user-level I/O isolation by providing multi-streaming to resist *write amplification*.

### 2.3 | 1-D K-means

K-means, an unsupervised learning method, is popular for its simplicity. In our work, we cluster the applications based on the length of the application's I/O phrase so a 1-D K-means algorithm can solve the problem[34] with $O(n^2 K)$. There is a more efficient but complex algorithm with $O(nK + n \log n)$[35]. These work are offline and can be used for batch applications.

For dynamic coming applications, we need an online or streaming method. There are some researches on online K-means. In[36], an online K-means is proposed under a performance guarantee. But almost all online methods consider the increased data not for dynamic data, in which new data will arrive and old data might be clear away. In[37], a clustering method over evolving stream data in an IOT scenario is proposed by capturing data distribution change. It motivates our distribution-based thresholding for updating the clustering.

## 3 | PRELIMINARIES AND MOTIVATIONS

In this section, we first present the platform architecture of Burst-Buffers deployed HPC. Then, we introduce the application execution model and probabilistic model. Finally, we describe the objectives of I/O scheduling and the motivations of our work based on some observations.

### 3.1 | Platform Architecture

A traditional HPC platform contains a large number of computing nodes and storage nodes to support large-scale scientific computing. The computing nodes are entirely identical with the same computing power and the same local network bandwidth. Many applications run on such a platform concurrently, and each occupies a fixed set of computing nodes arranged by an application scheduler.

In this work, the applications are placed on the computing nodes by an application scheduler ahead. When accessing the shared storage nodes for I/O operations, the applications will concurrently communicate with the shared file system that provides an aggregated bandwidth $B$. For the convenience of analysis, we assume one single file server as PFS. (In fact, there may be several file servers to provide the aggregated bandwidth.) To prevent I/O congestion, modern HPC architectures contain a Burst-Buffer system to absorb I/O bursts. We take into account this characteristic in our platform model. The Burst-Buffers is modeled with size $S$ in a pseudo-centralized way. The model's schematic view is shown in Figure 1.

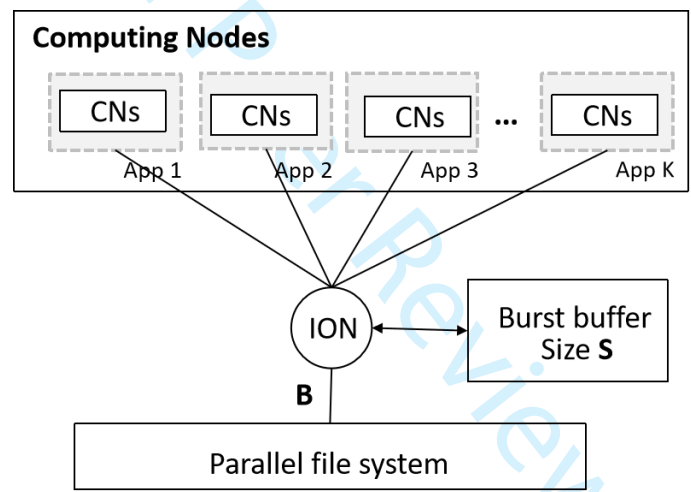

**FIGURE 1** Schematic model of Burst-Buffers equipped HPC

### 3.2 | Application Model

#### 3.2.1 | Application Execution Model

To model applications realistically, we consider the main I/O characteristics of applications running on modern HPC platforms summarized in [19], such as periodicity and burstiness. For studying the effect of Burst-Buffers, we focus on the write-intensive applications mainly, like the Check-pointing of large scientific simulation. The execution of this kind of application consists of two phases: computation and I/O. Applications alternative these two phrases and show apparent periodicity. Since many such applications run concurrently on the HPC platform and exhibit the characteristic of I/O burstiness, I/O congestion can be caused by sending excessive I/O requests to the bandwidth-constrained, shared I/O network.

To clarify, we illustrate an example of application execution in Figure 2. Three applications have different computational patterns. $App^1$ has a volume $W^1$ of computing work with a medium I/O load. $App^2$ has a volume $W^2$ of computing work with a high I/O load. $App^3$ has a volume $W^3$ of computing work with a lower I/O load. They all take the *best effort* strategy to access the shared PFS. When the load of I/O operations up to the maximum aggregate bandwidth of PFS, I/O congestion emerges as shown in Figure 2.

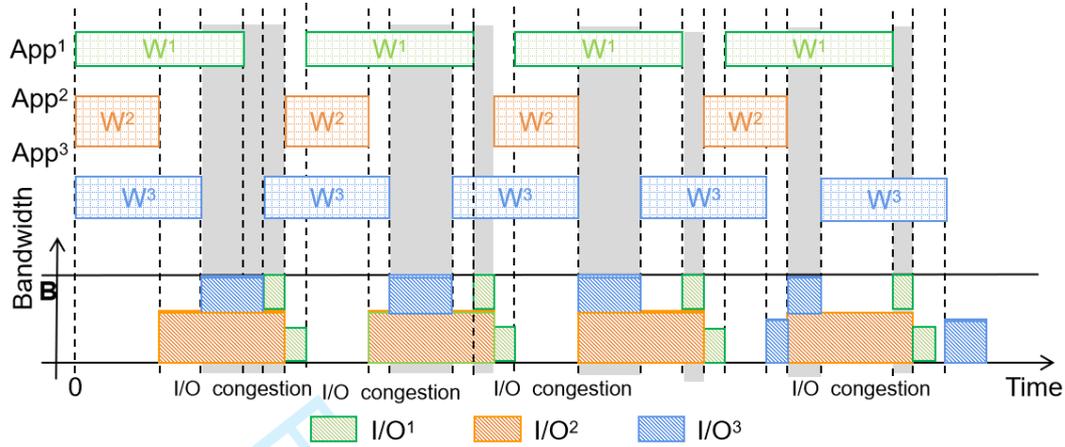

**FIGURE 2** Example for application execution with best effort I/O strategy

In fact, the applications are submitted to I/O scheduler in batches, sometimes sporadically as illustrated in section 4.1. The description here is a snapshot of applications' execution on the system. We assume there are $K_A$ applications running on the HPC platform. Application $A_i$ runs on $\beta_i$ computing nodes which is allocated by a batch job scheduler. There are $N_i$ instances for each application $A_i$ that alternates between computing phase and I/O phase with a quasi-periodic duration $T_i$. In the computing phase of an instance, there are $W_i$ unit time workload for computing. The maximal rate for transferring data is $B_i$, which is determined by $\beta_i$ computing nodes. Namely, $B_i \propto \beta_i$. In the I/O phase, there are a volume $IO_i$ of I/O workload. The actual I/O rate $b_i$ is the minimum between $B_i$ and the remainder of the PFS bandwidth. That is $b_i = \min\left(B_i, B - \sum_{i \neq j} b_j\right)$.

Due to the existence of Burst-Buffers, it is unnecessary to block or slow down some data transformations when congestion occurs on PFS. The Burst-Buffers can absorb the excess if the velocity of transferring data exceeds the bandwidth to PFS. If the load continues to exceed $B$, the burst buffer may be full, causing congestion again. To deal with this situation, I/O scheduling as a complementary way can reorder the I/O requests from different applications to mitigate this kind of I/O congestion. So it is essential to model the state of burst-buffer for determining the specific scheduling. In this work, we utilize the probabilistic model of application and the Markov-Chain to model the states of Burst-Buffers, which was introduced in our previous work[23].

### 3.2.2 | Application Probabilistic Model

As described above, application $A_i$ executes periodically with the period $T_i$ and the proportion of I/O phase is $P_i$, where $P_i = 1 - W_i/T_i$. So the expected duration of data transferring for application $A_i$ is $P_iT_i$ and then the duration of computing is $(1-P_i)T_i$. Without loss of generality, we can combine the above static description and this probability representation to get the execution time of computing phrase $W_i = (1-P_i)T_i$ and the I/O volume of I/O phrase $IO_i = P_iT_iB_i$. To verify the efficiency of our model, we set these parameters by realistic values shown in Table 1 that origins from the APEX report as same as the Aupy's work[2].

In order to analyse the states of Burst-Buffers, we discretize the continual time to obtain the probability of transferring data in a time unit for each application. The discretized time unit is set as the average value of the I/O time, $\frac{1}{K_A}\sum_i P_iT_i$, which is significantly smaller than the period but comparable to the length of short I/O operations. In a time unit, each application $A_i$ takes I/O operations with probability $P_i$. This assumption is crucial for estimating the instantaneous total load and building the Markov-Chain model.

### 3.2.3 | Application-Clustering based Probabilistic Model

In this work, the proposed approach separates all the applications into some clusters through 1-D K-means clustering on their I/O lengths. In Figure 3, we show an example that illustrates three clusters with different discretized time units. Each cluster, $C_i$, consists of some applications with similar I/O characteristics, which satisfies the consistency assumption described above.

| Workflow | EAP | LAP | Silverton | VPIC |
|---|---|---|---|---|
| Number of instances | 13 | 4 | 2 | 1 |
| $B_i$ rate (GB/s) | 160 | 80 | 160 | 160 |
| $T_i$ Period (s) | 5671 | 12682 | 15005 | 4483 |
| Checkpoint time (s) | 20 | 25 | 280 | 23.4 |
| $P_i(\times 10^{-3})$ | 3.51 | 1.97 | 18.7 | 5.11 |

**TABLE 1** Characteristics of the APEX applications data set.[2]

Their discretized time units are defined as the average of their I/O time, $\frac{1}{|C_i|}\sum_{i\in C_i} P_i T_i$. If we do not separate the applications with different I/O characteristic, the built probabilistic model might be useless to modeling the congestion of Burst-Buffers.

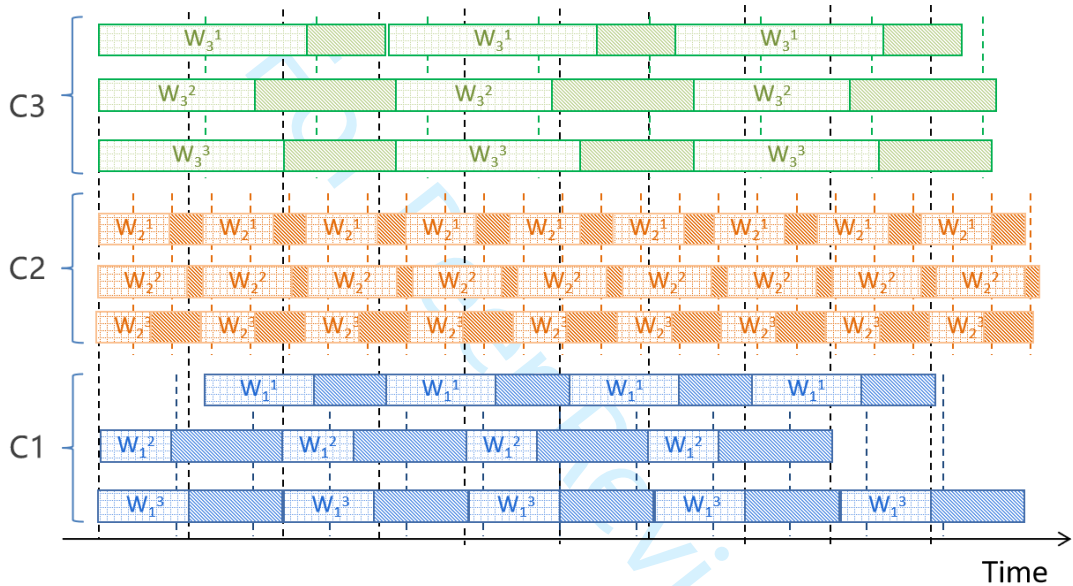

**FIGURE 3** Example for application clustering

## 3.3 | Objectives of I/O Scheduling

Our goal of I/O scheduling is to maximize system efficiency and minimize application dilation, same as in Gainaru's work[11]. We will use these objectives to guide the I/O scheduling.

First, we need to define the application efficiency for each application $A_i$ at time $t$.

$$\tilde{\rho}_i(t) = \frac{\sum_{i\leq n_i(t)} W_i}{t - r_i},$$

where $n_i(t) \leq N_i$ is the number of instances (each instance consists of a computing phrase and an I/O phrase) of $A_i$ that have been executed at time $t$, $r_i$ is the release time of $A_i$. The optimal application efficiency $\rho_i$ is the rate obtained in the dedicated mode. $\rho_i = \frac{W_i}{T_i}$ and $\tilde{\rho}_i(t) \leq \rho_i$.

Therefore, the two key objectives can be defined as follows:

- MaxSysEfficiency: to maximize the peak performance of the platform– that is the total performance of all processors in the platform.

$$\max \frac{1}{N} \sum_{i=1}^{K_A} \beta_i \tilde{\rho}_i(d_i) \quad (1)$$

where $d_i$ is the finish time of application $A_i$.

- MinDilation: to minimize the largest slowdown among applications.

$$\min \max_{i=1\ldots K_A} \frac{\rho_i(d_i)}{\tilde{\rho}_i(d_i)} \quad (2)$$

The rationale behind the *MinDilation* objectives is to provide fairness between all applications. It guides the scheduling to minimize the maximum of slowdowns to avoid starvation of applications.

## 3.4 | Motivations of This Work

When the spare space of burst-buffers is insufficient, the I/O performance might be degraded dramatically due to the *write amplification* of Burst-Buffers [24]. Therefore, the probabilistic I/O scheduling [22,23] utilizes the state of burst-buffers to mitigate I/O congestion.

However, the probabilistic I/O scheduling has the limitation of consistent I/O characteristics [23], which is often violated in practice. To eliminate this restriction, we aggregate the similar applications together by 1-D K-means clustering and update the clusters with an online method. In this way, the estimation of system workload can be more precise for accurate prediction of the congestion on Burst-Buffers. In the meantime, because a Burst-Buffer is divisible and randomly-accessible [8,14], it can be partitioned easily and adjusted dynamically.

As a consequence of the above observations, , we deploy a clustering algorithm to cluster the applications and an online method to update these clusters dynamically. So we can build an accurate prediction model for the state transition of Burst-Buffers. Finally, we apply our scheduler [23] to schedule periodic I/O operations without consistency limitation.

## 4 | DYNAMIC PROBABILISTIC I/O SCHEDULING

In this section, we describe our framework of dynamic probabilistic I/O scheduling via application clustering (DPSAC) in detail. We first introduce the overall structure and application execution process, and then show the function and design of each part in the framework.

## 4.1 | Framework Structure and Execution Process

The framework schedules I/Os of periodic applications with different I/O characteristics on Burst-Buffers equipped HPC. As depicted in Figure 4, the framework includes four main components: Aggregator, Partitioner, Updater, and I/O Scheduler. Aggregator deploys a 1-D K-means clustering algorithm to aggregate all applications in a batch into some clusters by their I/O phrase lengths and merges the clusters into the existing clusters on the system. Partitioner calculates the workload of each cluster based on the probabilistic model of application and partitions the Burst-Buffers proportionally. Finally, Updater updates the configuration when some events occur, allowing the framework to break the limitation of application length and enable applications to be joined and removed dynamically via thresholding or an online k-means algorithm. I/O Scheduler decides the order of I/O for all applications.

As an example illustrated in Figure 4, some applications $\{A_1, A_2, \cdots, A_i\}$ are submitted into HPC in a batch and the job scheduler on HPC allocates them into some computing nodes. The Aggregator of the I/O scheduling framework clusters them into three clusters, $C1, C2, C3$, where each cluster only includes the applications with similar I/O characteristics. Then Partitioner calculates the expected workload for each cluster to get the workloads, $L1, L2, L3$, and partitioned the Burst-Buffers into three parts proportionally. The I/O Scheduler schedules the I/Os from different applications in each cluster based on a optimizing strategy derived from the optimization objectives. When a new application, $A$, comes, the Updater inserts it into a cluster with similar I/O length via thresholding or an online K-means algorithm.

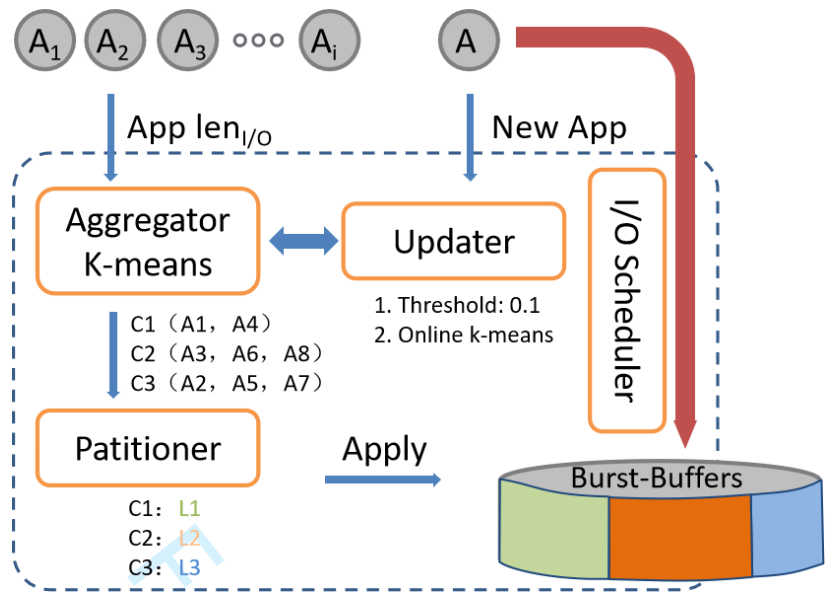

**FIGURE 4** Probabilistic I/O scheduling framework (DPSAC)

The framework DPSAC allows applications come in batches or individually, and some applications can be eliminated during the execution due to completion or forcibly termination. On HPC platform, all applications are submitted in batches, and the job scheduler allocates the computing resources to each application. The I/O resources (PFS and Burst-Buffers) are shared in the pseudo-centralized architecture. As the running progress, some applications finish and release some computing resources. The job scheduler continues to allocate the spare resources to submitted applications to improve system utilization. This running way causes that the applications come into the I/O scheduler in batches or individually. To show the execution process of the framework, we illustrate the sequence diagram of a schematic example in Figure 5.

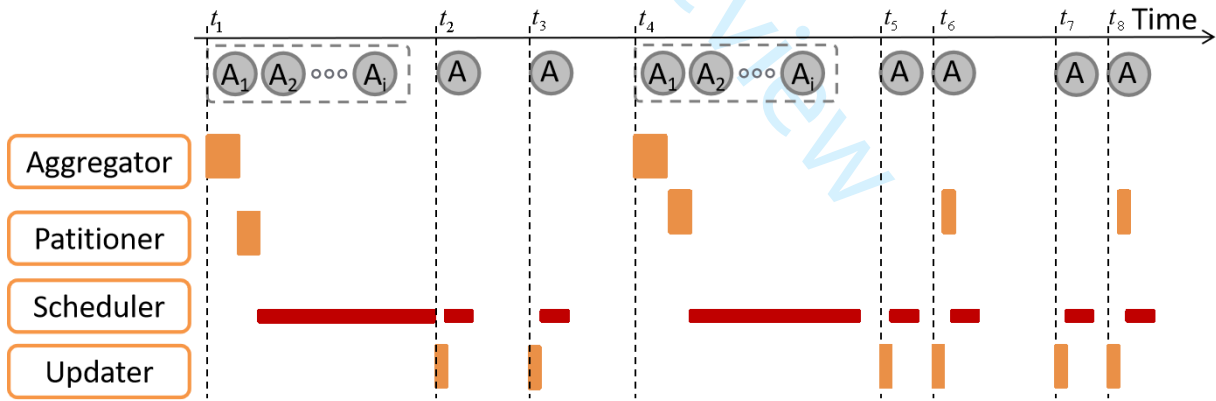

**FIGURE 5** Applications execution process of DPSAC

As depicted in Figure 5, the four components of the framework coordinately schedule the application's I/O. Specifically, at $t_1$, a group of applications comes in a batch. The Aggregator first cluster it by 1-D K-means algorithm and then Partitioner calculates the workload of each cluster and divides the burst buffers proportionally. Next, the Scheduler performs the I/O scheduling based on a optimizing strategy. At $t_2$ and $t_3$, a few applications join or exit. Updater updates the workload online without changing the clustering. At $t_4$, another group of applications comes. There is a choice on updating the clusters by either inserting each application to the closest existing cluster or clustering the whole batch and then merging the result into the existing clusters. We use a heuristic strategy to make the choice: insertion if the number of applications in the group is not greater than the average cluster size of the existing clusters; clustering and merging otherwise. The situation at $t_5$ and $t_7$ is similar to $t_2$ and $t_3$. Nevertheless, at $t_6$ or $t_8$ the added applications make an obvious change on the distribution of applications, so the Partitioner needs to update the partition of Burst-Buffers.

## 4.2 | Clustering Applications

For most HPC platforms, applications are scheduled in batches and some temporary ones may be submitted between two batches to make full use of idle resources. The Aggregator in our framework clusters the batch of applications first and then merges the generated clusters with the existing clusters.

For the current batch, there are $N_A$ applications, $S_A = \{A_1, \cdots, A_{N_A}\}$, as described in the application model section. The applications can be classified into some clusters based on their lengths of I/O phrase. The number of these clusters, $K_C$, is usually small e.g. $K_C \leq 5$. Typically, HPC platforms divide applications into small and large parts.[11] This is the reason that K-means is chosen rather than other clustering methods.

Since we cluster the applications based on the length of application's I/O phrase, a 1-D K-means clustering algorithm, known as Ckmeans.1d.dp[34], can solve the problem with $O(n^2 K)$. We deploy the dynamic programming algorithm due to its simplicity, though there is a more efficient algorithm with $O(nK + n \log n)$[35]. The sum of squared Euclidean distances to the mean of each cluster is chosen as our measure in clustering applications, instead of the cluster length (like some 1-D clustering), since the elements in each cluster should be centralized rather than distributed.

The 1-D K-means problem is defined as follows. Let $L_A = \{l_1, l_2, \cdots, l_{N_A}\}$ be an input array represented the length of application's I/O phrase from $S_A$, which has been sorted in non-descending order. Assign each element in the input array into K clusters so that the sum of squares of within-cluster distances, $wss$, is minimized.

$$\min \sum_{i=1}^{K} \sum_{l \in C_i} ||l - c_i||_2^2$$

where $l$ is an element from $L_A$ and $c_i$, $c_i = (1/|C_i|) \sum_{e \in C_i} e$, is the center of cluster $C_i$ that corresponds to a set $S_i$ of applications.

The dynamic programming algorithm constructs a matrix, $D$, with $N_A + 1$ rows and $K_C + 1$ columns to store the immediate results. Each entry, $D[i, m]$, in the matrix is the minimum $wss$ for clustering the first $i$ elements from $L_A$ into $m$ clusters. For a optimal clustering $D[i, m]$, we let $j$ be the index of first element in $m$-th cluster, and its first $m - 1$ clusters in this clustering correspond to $D[j - 1, m - 1]$. So the recurrence equation for dynamic programming can be expressed as:

$$D[i, m] = \min_{m \leq j \leq i} \{D[j-1, m-1] + d(l_j, \cdots, l_i)\},$$
$$1 \leq i \leq N_A, 1 \leq m \leq K_C$$

where $d(l_j, \cdots, l_i)$ is the sum of squared distances from $l_j, \cdots, l_i$ to their mean, which can be calculated within constant time[34], and the 0-th row and 0-th column of the matrix is initialized to zero, $D[0, :] = 0, D[:, 0] = 0$.

We adopt the original dynamic programming algorithm to meet our need for the unknown number of clusters, $K$. Because $wss$ for all $K$ is provided in matrix $D$, the standard Elbow method can be used to determine which $K$ is most appropriate. An additional matrix, $C_I$, is introduced to store the index of the first element of the last cluster in the optimal clustering. The specific description is shown in Algorithm 1. The overall time complexity is $O(n^2 K)$ and the space complexity is $O(nK)$.

After clustering on the current batch, we need to merge the new generated clusters with the existing clusters. If the number of the new clusters is smaller than the number of existing clusters, we merge each new cluster into the closest existing cluster. Otherwise, we merge each existing cluster into the closest new cluster.

## 4.3 | Partitioning the Burst-Buffers

The Burst-Buffers are easy to be partition because it has divisibility and access-randomness, which is decided by the characteristics of the underlying SSDs. We just need to calculate the expected I/O workload for each cluster obtained by Aggregator and then proportionally partition the Burst-Buffers.

Based on the probabilistic model of applications mentioned in Section 3.2.2, the expected I/O load, $EIO_k$, for cluster $S_k$ is given by $EIO_k = \sum_{i \in S_k} P_i B_i$ for every time unit. All I/O workload described in this paper is defined in the term of time unit. The total expected I/O load of all clusters, $TEIO = \sum_{k \leq K} EIO_k = \sum_i P_i B_i$, should be smaller than the bandwidth of PFS, ie.

**Algorithm 1** 1-D K-means with dynamic programming

**Input:** Applications $S_A = \{A_1, \cdots, A_{N_A}\}$
**Output:** Clusters $S = \{S_1, \cdots, S_K\}$
1: Obtain $L_A = \{l_1, l_2, \cdots, l_{N_A}\}$ by sorting the the I/O phrase's length of all applications from $S_A$ in non-descending
2: Declare two $N_A + 1$ by $K_C + 1$ matrix $D$ and $C_I$
3: **for** $0 \leq i \leq N_A, 0 \leq m \leq K_C$ **do**
4:     **if** $i == 0$ or $m == 0$ **then**
5:         $D[i, m] = 0$ and $C_I[i, m] = 0$
6:     **else**
7:         $D[i, m] = \min_{m \leq j \leq i}\{D[j - 1, m - 1] + d(l_j, \cdots, l_i)\}$
8:         $C_I[i, m] = j$
9:     **end if**
10: **end for**
11: Get the appropriate number of clusters by Elbow method
12: Obtain $C = \{C_1, \cdots, C_K\}$ by searching $C_I[N_A + 1, K]$
13: Return the application clusters $S$ corresponding to $C$

$TEIO < B$, which is the real situation observed on almost all data centers[8]. If the total expected I/O load is greater than $B$, I/O transferring will be delayed by some delay mechanism.

We then partition the Burst-Buffers based on $EIO_k$ of each cluster into $K$ parts proportionally. For the good quality of Burst-Buffers, we assign an usage for each cluster, $U_k \cdot S$, where $U_k = \frac{EIO_k}{TEIO}$, where S is the capacity of Burst-Buffers.

### 4.4 | I/O Scheduling

In this section, we first obtain the instantaneous I/O load distribution, $Dist$, by a method introduced in[2], for each cluster accurately based on the probabilistic model of application and the clustering result of Aggregator. Then, the load transition matrix and the transition-to-full probability on a partition of the Burst-Buffers are established to capture the state change of buffers based on the Markov-Chain model[23]. Finally, probabilistic I/O scheduling for each partition is introduced briefly.

#### 4.4.1 | Instantaneous I/O Load Distribution

Since the clustering algorithm already clusters the applications by I/O characteristic, the length of each application's I/O phrase has a similar size in the same cluster so that a more accurate I/O load distribution can be built. In practical, the real instantaneous I/O load should be given by $InstantIO(t) = \sum_{i \in A(t)} B_i$, where $A(t)$ is the set of applications performing I/O at time $t$. But this runtime information is useless for on-time decision-making. So we build the instantaneous I/O load based on the probabilistic model of application.

For a cluster $S_k$, an application $A_i \in S_k$ will perform I/O transferring with probability $P_i$ in each time unit. We define a 0-1 random variable $X_i$ to indicate whether $A_i$ is sending I/O and all $X_i$ are mutually independent. The instantaneous I/O load can be defined as $X = \sum_{A_i \in S_k} B_i X_i$ based on the application probabilistic model.

To obtain the instantaneous I/O load distribution for each cluster, we apply the iterative algorithm with time complexity $N \sum_i B_i < BN^2$ [22], which adds applications one after one and updates the distribution. $Dist(k)$ denotes the probability that instantaneous I/O load is $k$. This detailed algorithm is shown in Algorithm 2. This iterative algorithm is suitable for the dynamic change of application and be extended to an online version.

#### 4.4.2 | Transition-to-full Probability

Due to *write amplification* of SSD[24], the performance of Burst-Buffers will rapidly decrease with its spare space exhausting. We take the state information of Burst-Buffers into account when scheduling I/O. The state of the buffers is modeled by a Markov chain. Based on the application's instantaneous I/O load distribution for each cluster, the Load state transition matrix can be obtained.

---

**Algorithm 2** Instantaneous I/O load distribution $Dist$
**Input:** Applications $A_i(B_i, P_i)$
**Output:** $Dist$, such that $Dist(k) = Pr(\sum_i B_i X_i = k)$
    $Dist$ is a vector with size $\sum_i B_i$
1: $Dist(0) = 1$ and $Dist(k) = 0$ for each $k \neq 0$.
2: **for** each $A_i$ **do**
3:     **for** each $k$ **do**
4:         $Dist(k) = (1 - P_i)Dist(k) + P_i Dist(k - B_i)$ if $k \geq B_i$;
5:         $Dist(k) = (1 - P_i)Dist(k)$ if $k < B_i$;
6:     **end for**
7: **end for**

---

Let $Y_j, 0 \leq j \leq U_k \cdot S$ be each buffer state for the cluster $S_k$. Any transition probability from $Y_m$ to $Y_n$ can be estimated as $P_{m,n} = Dist(n - m + U_k \cdot B)$ when the buffers are not empty or full. To mitigate I/O congestion, our focus is on the probability of transiting to the full state $Y_{U_k}$, given by $P_{j, U_k \cdot S} = Dist(k \geq U_k \cdot S - j + U_k \cdot B)$. When the I/O load is greater than the sum of the spare space of Burst-Buffers for current state $j$ and the bandwidth of PFS, I/O congestion on Burst-Buffers happens.

### 4.4.3 | Probabilistic I/O Scheduling

To clarify the description, we introduce the basic probabilistic I/O scheduling, and more sophisticated variations can refer to our previous work in [22,23]. This basic approach is used as a combination of common I/O scheduling for different optimizing objectives and Burst-Buffering to take the state of Burst-Buffers into account.

First, the algorithm obtains a sequence of applications with priority based on a heuristic strategy for some optimizing objective. The I/O scheduling problem in our situation is NP-complete, and some heuristic strategies are described in Guinaru's work [11]. *MinDilation* strategy favors the applications with low values of $\frac{\tilde{\rho}_i(t)}{\rho_i(t)}$ to reduce the application dilation; *MaxSysEff* strategy favors the applications with high $\beta_i \frac{\rho_i(t)}{\tilde{\rho}_i(t)}$; *MinMax-$\gamma$* strategy is for a balance between *MinDilation* and *MaxSysEff*.

Then, the priority sequence is modified by combining the state of Burst-Buffers to reduce the I/O congestion. Specifically, we obtain the usage of each partition of Burst-Buffers for each cluster and then calculate the transition-to-full probability to determine whether transferring I/O for each application.

The particular algorithm is described in Algorithm 3. At each unit time, the bandwidth allocation for each active application considers the load (used space) of cluster partition and the instantaneous I/O load of all applications in the cluster. The heuristic strategy is used to optimize the objectives. When the instantaneous load is less than $B \cdot U_k$, the Burst-Buffers can be empty partially. Otherwise, partial I/O load will be transferred to buffers. To decrease the I/O load, we select the application transferred with probability [23] and apply this kind of probabilistic I/O scheduling to each cluster of applications.

## 4.5 | Dynamic Updating the Clusters

In this section, we propose two kinds of methods for dynamically updating the existing clusters in the system when some applications arrive or terminate. In the meantime, Updating the partition of Burst-Buffers and the I/O load distribution is also introduced.

### 4.5.1 | Updating with Thresholding

For thresholding, we provide a simple heuristic approach and a distribution-based approach based on the concept of data drift [37]. The idea behind these approaches is to capture the change of data distribution over time.

#### a) Simple Thresholding

For the applications in a batch, the Aggregator can generate clusters, $C = \{C_1, \cdots, C_K\}$, on the length of the application's I/O phrase by clustering and merging. Each cluster $C_i$ has a centroid $c_i$. Thresholding sets a bound, 0.1, for the cumulative change of

---

**Algorithm 3** Probabilistic I/O scheduling for clusters

**Input:** A set of active application $A_i(W_i, IO_i, B_i, r_i)$, PFS bandwidth $B$, Burst buffer size $S$, the load of cluster partition $L_k$, usage share $U_k$ for each cluster, clusters $S_k$

**Output:** The bandwidth allocation for each active application

1: **for** each cluster $S_k$ **do**
2:    **for** each $A_i$ in $S_k$ **do**
3:       Compute the current application efficiency $\tilde{\rho}_i(t)$
4:    **end for**
5:    Based on a heuristic strategy, compute the priority of applications to get the order $(k_1, ..., k_l)$
6:    **if** $\sum_{i=1}^{l} B_{k_i} \leq B \cdot U_k$ **then**
7:       Allocate the bandwidth $B_{k_i}$ for application $A_{k_i}$
8:       Empty the burst buffer with $B \cdot U_k - \sum_{i=1}^{l} B_{k_i}$
9:    **else**
10:       Obtain the I/O load distribution $Dist$ by Alg. 2
11:       The probability $P_{full} = Dist(k \geq (S + B) \cdot U_k - L_k)$
12:       Get $m$ so that $\sum_{i=1}^{m-1} B_{k_i} < B \cdot U_k$ and $\sum_{i=1}^{m} B_{k_i} \geq B \cdot U_k$
13:       Allocate the bandwidth $B_{k_i}$ for first $m - 1$ applications transferring
14:       Allocate the bandwidth $B_{k_i}$ for application $A_{k_i}, i \geq m$ with a probability $(P_{full})^{i-m}(1 - P_{full})$
15:       **if** $\sum_{\text{selected } A_{k_i}} B_{k_i} - L_k - B \cdot U_k < S \cdot U_k$ **then**
16:          Transfer with bandwidth $B \cdot U_k$ to PFS and with $\sum_{\text{selected } A_{k_i}} B_{k_i} - B \cdot U_k$ to buffer
17:       **else**
18:          Allocate the bandwidth $B_{k_i}$ for first $m - 1$ applications and $B \cdot U_k - \sum_{i=1}^{m-1} B_{k_i}$ for $m$th application
19:       **end if**
20:    **end if**
21: **end for**

---

the centroid incurred by the change of applications. We believe 0.1 is significant here and this bound can be adjusted depending on the actual situation.

When an application $A$ with I/O length $l$ comes, it is inserted into the closest cluster $C_i$. Then, the cumulative centroid becomes $\hat{c}_i \leftarrow \frac{\hat{c}_i \cdot |C_i| + l}{|C_i| + 1}$. When the cumulative change $\frac{\hat{c}_i - c_i}{\max\{C_i\} - \min\{C_i\}} \geq 0.1$ and the number of elements in this cluster exceeds $K_C$, we split the cluster at the biggest interval between two consecutive elements. The Partitioner then modifies the usage quotas $U_k$ for each cluster. The instantaneous I/O load distribution generated by the iterative Algorithm 2 is also updated iteratively for the newly coming applications. For other cluster updating approaches, the updating on Partitioner and I/O scheduler is the same.

When an application exits, we remove it from the corresponding cluster. The workload on the corresponding partition is decreased. We do not need to update our framework's parameters since removing applications does not increase the I/O workload, and the parameters will be corrected when a new application is added.

### b) Distribution-based Thresholding

In order to capture the dynamic change of applications, we track the change of distribution over all lengths of the application's I/O phrase. This distribution-based method is motivated by the concept of data drift to cluster over the evolving stream data[37]. Then we use *Jensen-Shannon divergence*[38] to describe the difference between two distributions for its symmetry compared with *Kullback-Leibler divergence*.

Like the simple thresholding, we generate the clustering for applications arriving in batches by Aggregator, and then create a distribution, $p$, over the length of application's I/O phrase, which records the count of applications in each interval.

When applications running on the HPC platform come, we just insert it into the closest cluster and modify the distribution accordingly. After $K_C$ changes happened, we use *Jensen-Shannon divergence* to measure the difference between the new distribution $q$ and the original distribution $p$ as follows.

$$D_{JS}(p||q) = \frac{1}{2}D_{KL}(p||\frac{p+q}{2}) + \frac{1}{2}D_{KL}(q||\frac{p+q}{2}),$$

$$\text{where} \quad D_{KL}(p||q) = \int_x p(x) \ln \frac{p(x)}{q(x)} dx$$

When the distribution change $D_{JS}(p||q) \geq 0.1$, we need to adjust the clustering. Suppose the existing clusters, $\{C_1, C_2, \cdots, C_K\}$, have the corresponding centroid $c_i$. We calculate $\frac{|p(c_i)-q(c_i)|}{p(c_i)}$ (that represents the distribution change within a cluster) for each cluster and select the cluster $C_o$ corresponding the maximum. We split the cluster $C_o$ at the biggest interval between two consecutive elements in $C_o$ and then update the distribution $p$ within the cluster $C_o$. The other updating on Partitioner and I/O scheduler is the same as the simple thresholding. The time complexity of comparing two distributions depends on the size of the interval.

### 4.5.2 | Updating with Online K-means

Online K-means clustering gets rid of the threshold dependency and maintains a set of cluster centers to capture the data change. We adopt the online K-means Algorithm from [36] to cope with the application dynamic changes. The original algorithm like many other online or streaming K-means algorithms was designed for the addition of new points, not for removing the existing ones.

In our online updating algorithm, we first create the original clustering with $k$ clusters for the applications in batches by the Aggregator. When a new application arrives, it decides whether to add the new as a center into the clusters $C$ with a probability proportioning to the distance between it with the clusters, $D(v, C) = \min_{c \in C} ||v - c||_2$. The detailed algorithm is shown in Algorithm 4. The *yield* command updates the center for the new application $v$. If $v$ is added into $C$, then the original center closest to $v$ is removed; otherwise, the original closest center remains.

To remove an application, we use a pessimistic way to delete the application from its cluster directly. This online algorithm is in favor of the application addition. However, this algorithm can give a quantity guarantee and more details can refer to [36].

---

**Algorithm 4** Online K-means updating

**Input:** All the application changes $V$, and $k$
**Output:** The set of cluster centers $C = \{c_1, \cdots, c_k\}$
1: Initialize $C$ with Aggregator
2: $w^* = \min_{v,v' \in C} ||v - v'||^2 / 2$
3: $r = 1; q_1 = 0; f_1 = w^*/k$
4: **for** $v \in$ the remainder of $V$ **do**
5:     $n = n + 1$
6:     **if** $v$ is new arrival application **then**
7:         with probability $p = \min(D^2(v, C)/f_r, 1)$
8:             $C = C \cup \{v\}; q_r = q_r + 1$
9:         **if** $q_r \geq 3k(1 + log(n))$ **then**
10:             $r = r + 1; q_r = 0; f_r = 2 \cdot f_{r-1}$
11:         **end if**
12:         **yield:** the center $c = arg \min_{c \in C} ||v - c||^2$
13:     **else**
14:         **if** $v \in C$ and no other application with the center $v$ **then**
15:             Remove $v$ from $C$
16:         **end if**
17:     **end if**
18: **end for**

---

In this section, we provide two kinds of updating methods, including two thresholdings, for their respective advantages. Simple thresholding is easy and effective with low computation complexity. Distribution-based thresholding can capture the entire change over distribution but lose some performance to compare two distributions. To eliminate the limitation of threshold with more computation, updating with an online K-means provides an approach to obtain a result quality guarantee, $O(W^* \log n)$, where $W^*$ denotes the value of the optimal solution and $n$ denotes the number of online added applications.

## 5 | SIMULATION RESULTS

In this section, we design some simulation experiments to evaluate the performance of our proposed framework, DPSAC. The experiments are conducted on applications that have diverse I/O characteristics under different I/O congestion settings. The first set of simulations compares the performance of clustering-based framework with Markov-Chain based scheduling without the limitation of application size. The second set of simulations evaluates the framework performance for different cluster updating strategies to demonstrate the benefits of each.

### 5.1 | System Configuration

To simulate the real Burst-Buffers equipped platform effectively, we set the related system parameters by referring to the experiment in [2,39], which comes from the Intrepid Blue Gene/P supercomputer in Argonne National Laboratory, US. The Bandwidth of PFS $B$ is set as 100GB/s. The burst buffer size is set as 300GB. The peak bandwidth for each node is 1 GB/s, which is unimportant in our work because we are just concerned about the total bandwidth of applications.

All the simulations are based on a discrete event simulator introduced by [2], which maintains an event queue based on timestamp. The events in the queue represent the computation or I/O behavior from some applications. The simulator mimics the running of a real HPC platform.

### 5.2 | Experiments for Different I/O Congestion

With different combinations of the applications described in Table 1, we construct different I/O congestion scenarios to test the performance of clustering-based scheduling (DPSAC), basic I/O scheduling (BIOS)[11] and basic Markov-Chain based I/O scheduling (MCIOS)[22]. To obtain the optimizing objectives, we select the scheduling strategy as *MinMax*-0.5.

#### 5.2.1 | Application configuration

In order to simulate more real situations under I/O congestion, we generate more applications by modifying the original real application in Table 1. We select EAP, LAP, VPIC as small applications, scale EAP and VPIC by 10 times together with Silverton as the group of big applications, and scale EAP and LAP by 5 times, Silverton by 0.5 times to make up the medium size group. Here, size refers to the length of the application's I/O phrase.

To obtain the different degrees of I/O congestion, we combine the different numbers of applications. The detailed setting is shown in Table 2. The bold number in the table denotes the number of the respective application. The blank cell denotes no respective application. From set #1 to #10, the number of application LAP is decreased to reduce I/O contention caused by concurrent I/O transferring. And some medium and big applications are included in the set for the diversity.

#### 5.2.2 | Results and Analysis

We conduct the simulation for each set of applications and obtain the system efficiency and application dilation defined in 3.3. The test for each set is repeated 5 times and the average is shown on Figure 6. As the results shown, the system efficiency increases gradually and the application dilation decreases with the reduction of I/O congestion.

For system efficiency, at set #1, I/O congestion is the most problematic. None of the methods reach 30%. DPSAC gets nearly the same performance as MCIOS, but both achieved twice the performance of BIOS. The reason is that these two methods are able to use the state of Burst-Buffers to resist the degradation of I/O performance. Increasing the diversity of applications from set #2 to #5, our proposed DPSAC becomes better than MCIOS and BIOS because it clusters different applications for more accurate scheduling. Specifically, DPSAC outperformed MCIOS 16.4% at set #5. The system efficiency even reached 93%. From set #6 to #10, I/O congestion disappears gradually, resulting in all three methods progressively achieving the best performance.

For application dilation, at set #1 BIOS causes almost twice the delay due to I/O congestion. MCIOS and our DPSAC obtain nearly a 12% performance improvement using probabilistic scheduling. DPSAC suffers a bit of a performance penalty due to the extra computation for application clustering. From set #2 to set #5, DPSAC shows less dilation than MCIOS and BIOS as I/O congestion decreases and application diversity increases. As I/O congestion disappears from sets #6 to #10, application dilation gradually disappears.

| Set # | 1 | 2 | 3 | 4 | 5 | 6 | 7 | 8 | 9 | 10 |
|---|---|---|---|---|---|---|---|---|---|---|
| EAP | | | | | 1 | | | | 1 | |
| LAP | 10 | 8 | 6 | 4 | 2 | 3 | 2 | 2 | | 1 |
| VPIC | | | | | | | 1 | | 1 | |
| EAP5 | | | | 1 | | | 1 | | | 1 |
| LAP5 | 2 | 1 | 1 | 1 | 1 | | | | | |
| Silverton0.5 | | | | 1 | 1 | 1 | | 2 | | 1 |
| EAP10 | | | 1 | 1 | | | 1 | | | |
| VPIC10 | | 1 | | 1 | 1 | | | 1 | | |
| Silverton | 1 | | | 1 | 1 | 1 | | | 1 | |

**TABLE 2** Application combinations for different I/O congestion

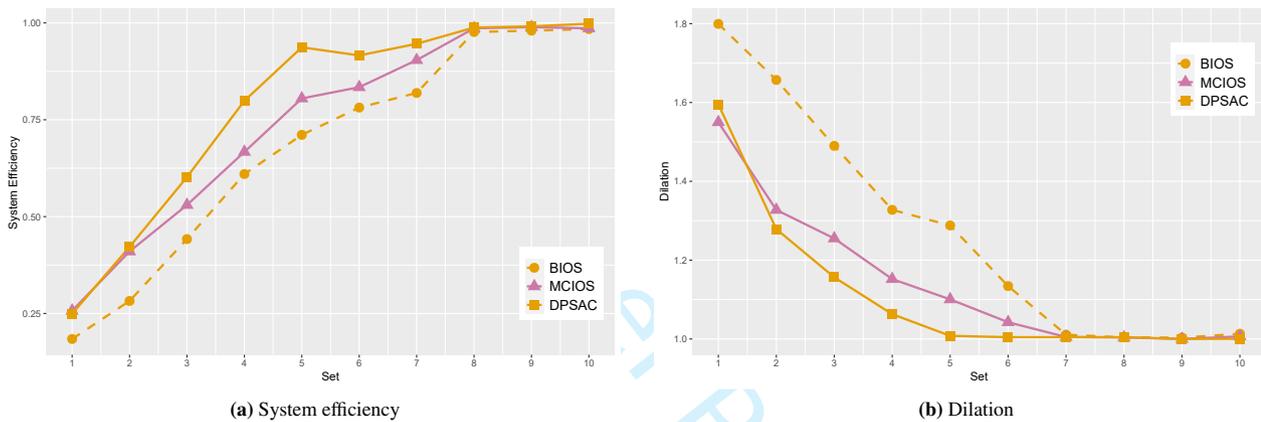

(a) System efficiency  (b) Dilation

**FIGURE 6** Performance of different scheduling for different I/O congestion

## 5.3 | Experiments for Different Updating Strategies

In order to demonstrate the ability of our proposed framework to adopt the dynamic changes in application workload, we need to set different sizes for each application and submit a variety of combinations of applications. We compare the three different updating strategies on these dynamic settings.

### 5.3.1 | Application configuration

In the dynamic running setting, applications are submitted in batches, and some maybe join in or exit. We use set #5 in Table 2 as a batch of applications since DPSAC obtains good clustering performance due to the diversity of applications, and specify each application a number of instances shown in Table 3. To simulate different types of application joining, we periodically join small (EAP), medium (LAP5), and big (Silverton) applications with different periods and different numbers of instances. See Table 4 for details.

| Application | # of applications | # of instances |
|---|---|---|
| EAP | 1 | **13** |
| LAP | 2 | **4** |
| LAP5 | 1 | **2** |
| Silverton0.5 | 1 | **2** |
| VPIC10 | 1 | **1** |
| Silverton | 1 | **1** |

**TABLE 3** Application in a batch

| Application | period (s) | # of instances |
|---|---|---|
| EAP | **6***5671 | 5 |
| LAP5 | **3***12682 | 2 |
| Silverton | **2***15005 | 1 |

**TABLE 4** Application for periodically joining

### 5.3.2 | Results and Analysis

We conduct the simulation for these application settings to test the performance of different updating strategies for different incoming applications. The updating strategies include Simple Thresholding (ST), Distribution-based Thresholding (DT) and Online K-means (OK). Each test is executed 5 times repeatedly for every kind of application.

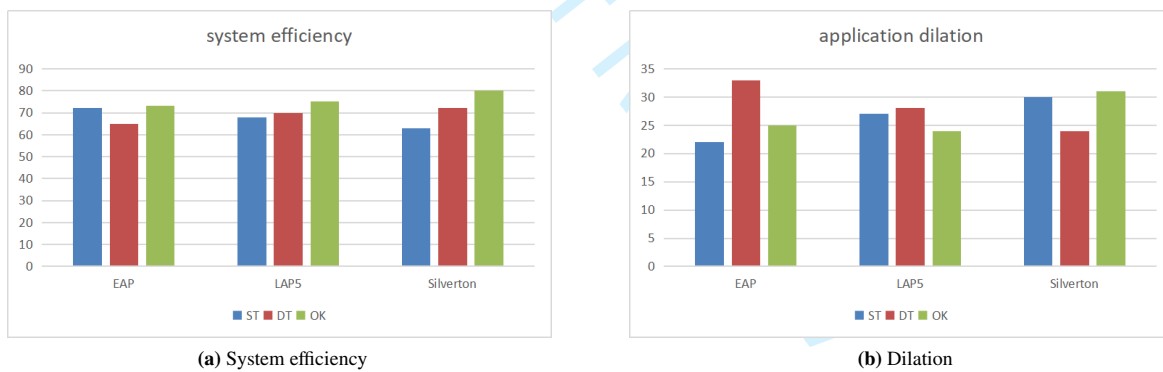

(a) System efficiency    (b) Dilation

**FIGURE 7** Performance for different incoming applications under different updating strategies

In the results shown in Figure 7, we note that the online K-means method produces a relatively better performance on system efficiency. That's because it guarantees application clustering performance. Simple Thresholding is good for small applications because it has a low computational complexity. Distribution-based Thresholding is better for big applications since it captures the overall changes in application distribution. For application dilation, Distribution-based Thresholding causes considerable application dilation for small applications due to the frequent computation involved in comparing distributions. Simple Thresholding shows less dilation since it requires less computation. Comparatively, Online K-means do not cause noticeable dilation.

## 6 | CONCLUSION

In this paper, we studied I/O scheduling on limited-size Burst-Buffers deployed HPC platforms for periodic applications. To overcome the limitation of the existing Markov-Chain based probabilistic I/O scheduling method that requires a consistent length of I/O operations and long execution time of applications, we proposed a generic framework of dynamic probabilistic I/O scheduling based on application clustering to decide I/O transferring online accurately.

The proposed framework contains four components: Aggregator clusters all applications based on the length of application's I/O operations through a 1-D K-means algorithm in polynomial time; Partitioner assigns usage share for each cluster to partition the Burst-Buffers based on the expected I/O workload obtained by the application's probabilistic model; Scheduler performs online I/O scheduling which considers the usage state of each partition; Updater supports the dynamic change of applications applying a updating strategy, thresholding or online K-means. The simulation experiment results show the superiority of the proposed clustering-based framework compared to the existing Markov-Chain method for a more generic situation. In the future, we will develop more efficient online clustering to handle dynamic data and take more real data characteristics into account.

## ACKNOWLEDGMENT

This work is supported by Key-Area Research and Development Plan of Guangdong Province #2020B010164003 and National Key Research and Development Plan's Key Special Program on High performance computing of China #2017YFB0203201.